# Selection and orientation of different particles in single particle imaging

Miklós Tegze* and Gábor Bortel

*Institute for Solid State Physics and Optics, Wigner Research Centre for Physics, Hungarian Academy of Sciences, H-1525 Budapest, P. O. Box 49, Hungary*

**Abstract.** The short pulses of X-ray free electron lasers can produce diffraction patterns with structural information before radiation damage destroys the particle. The particles are injected into the beam in random orientations and they should be identical. However, in real experimental conditions it is not always possible to have identical particles. In this paper we show that the correlation maximization method, developed earlier, is able to select identical particles from a mixture and find their orientations simultaneously.

*Corresponding author. e-mail: tegze.miklos@wigner.mta.hu, tel: +36-1-3922222, fax:+36-1-3922219

## Introduction

Determination of the structure of small particles by diffraction of short X-ray pulses before radiation damage destroys the particle was suggested first by Solem (1986). Neutze et al. (2000) predicted that high-intensity ultrashort pulses from X-ray free electron lasers (XFELs) could provide a new approach to structural determinations with X-rays. Model calculations have shown that sufficient information might be collected before the sample is ionized and explodes due to Coulomb forces (Huldt et al., 2003; Jurek et al., 2004a, 2004b; Jurek and Faigel, 2008; Hau-Riege et al., 2005; Lorenz et al., 2012). Several XFELs were built since then and most of the technical problems were overcome. In recent publications (Miao et al., 1999; Chapman et al., 2006; Seibert et al., 2011; Martin et al., 2011, 2012; Loh et al., 2012) successful 2D reconstructions of particles from experimental data were presented. Diffraction of XFEL beam on nanocrystals demonstrated that 3D atomic resolution structure of a protein can be determined from such experiments (Chapman et al., 2011; Boutet et al., 2012). However, experimental atomic resolution 3D structures of non-crystalline particles have not been published yet.

In a single particle diffraction experiment particles are injected into the XFEL beam. When an X-ray pulse hits a particle, a diffraction image is recorded on a 2D detector placed downstream. Since the total number of photons diffracted on a single particle and collected by the detector is rather small (in the order of a few thousands for a large biomolecule), it is necessary to record many diffraction patterns on identical particles to obtain detailed structural information. The particles are injected into the beam usually in random orientation. In order to assemble a consistent 3D intensity distribution, the relative orientation of the particles need be determined for all patterns. Several methods were developed recently to obtain the relative orientations from the patterns themselves. Some of the methods use the information in a common line or arc along the intersection of pairs of patterns (Shneerson et al., 2008; Bortel and Tegze, 2011). Other methods use more sophisticated algorithms like generative topographic mapping (GTM, Fung et al., 2009; Moths and Ourmazd, 2011) or expectation maximization (EMC, Loh and Elser, 2009, Loh et al., 2010) to solve the orientation problem. There are also suggestions to determine the orientation from independent experimental data, the angular distribution of the particle fragments after the Coulomb explosion (Jurek and Faigel, 2013). In this paper we use the simple and efficient method based on correlation maximization (CM), developed earlier by the present authors (Tegze and Bortel, 2012). This method has the advantage that besides being much faster even for relatively small objects, the computational cost also scales favourably with the size of the particle (Tegze and Bortel, 2012). Once the orientations of the particles are found and a consistent 3D intensity distribution is constructed from the diffraction patterns, the structure can be solved using one of several phase retrieval methods (Fienup, 1982; Miao et al. 2001; Marchesini et al., 2003; Oszlányi and Sütő, 2004, 2005; Martin et al., 2012).

In real experimental conditions it is not always possible to have identical particles. Biological objects are not easy to separate, large protein molecules may have several conformations, or contamination may introduce different objects into the beam. If the particles are not identical, orientation methods may not work and construction of a consistent 3D dataset could be difficult. Schwander et al. (2010) has shown that different molecular conformations can be separated by the generative topographic mapping (GTM) method. Here we show that the computationally more efficient CM method is also capable to select identical objects from a mixture. We study three cases. In the first, we mix patterns of lysozyme and cytochrome molecules. These molecules have nearly the same size, but their structure is completely different. In the second, lysozyme and its complex with arginine (Arg) are mixed. The presence of arginine changes the structure of the lysozyme molecule only slightly. In this case we wanted to test whether these small changes in the structure, together with the scattering contribution of the relatively small arginine are sufficient to distinguish between the molecules. Finally, we attempt to select the components from a mixture of all three kinds of molecules.

**Methods**

*Measurement simulation*

Synthetic diffraction patterns were calculated for randomly oriented molecules as described in our earlier work (Tegze and Bortel, 2012). The structures of the cytochrome, lysozyme and Arg-lysozyme complex molecules (Fig. 1) were taken from the Protein Data Bank (PDB IDs: 2XL6, 3LZT, and 3AGI, respectively). Poisson noise was introduced according to realistic experimental conditions. The most important parameters of the diffraction patterns are shown in Table 1.

*Orientation*

We used the CM method to orient the diffraction patterns. It is a much simplified variant of the EMC method of Loh and Elser (2009), where the expectation maximization step is replaced by a simple search for the orientation with the highest correlation. The algorithm was described earlier in detail (Tegze and Bortel, 2012), here we give only an outline. The diffraction patterns represent spherical sections (parts of the Ewald-sphere) of a 3D intensity distribution in reciprocal space. The centres of the patterns are always at the origin of the reciprocal space, but their orientations are unknown. The essential part of the algorithm is the comparison of the measured (in this case simulated) diffraction patterns to a 3D intensity distribution in different orientations (defined on a grid, see Table 1 and Tegze and Bortel, 2012) and finding the best fitting orientation for each pattern. Then these orientations are used to construct a new 3D distribution from the measured intensities. Starting from a random distribution, convergence is usually reached after a few iterations.

The similarity between the measured patterns and cuts from the 3D distribution is expressed in terms of the Pearson correlation (Rodgers and Nicewander, 1988; Tegze and Bortel, 2012). We found, that convergence is faster if at beginning of the iteration process when correlations are low, we use only a certain fraction of the patterns (the ones with a correlation higher than a certain limit) to build the new intensity distribution. We used the following condition to select the patterns contributing to the new distribution:

$$c_m^{max} > c_{limit},$$
where $c_{limit}=\min(\text{median}(\{c_i^{max}\}), c_{limit}^0)$.

Here $c_m^{max}$ is correlation of pattern *m* in its best-fitting orientation. In other words, the best-fitting half of the patterns and patterns with correlations above a certain limit are used to construct the new intensity distribution. At the beginning of the iteration process all correlations are low, so $\text{median}(\{c_i^{max}\}) < c_{limit}^0$, therefore only half of the patterns contribute to the new distribution. When the correlations increase, more and more correlations are above $c_{limit}^0$. Eventually, when reaching convergence, all diffraction patterns are included if $c_{limit}^0$ is chosen properly. The chosen value of $c_{limit}^0$ should be slightly below the expected correlation values between the noisy measurement patterns and the perfect (almost noiseless) solution of the 3D intensity distribution. This value can be estimated from the noise (number of counts) in the diffraction pattern or found from the actual distribution of the $c_m^{max}$ values after convergence.

*Particle selection*

The same or a similar rule can be used to select diffraction patterns of one kind of particle out of a mixture. The mixture may contain two or more kind of particles in any distribution. During iteration, those patterns will tend to contribute to the 3D intensity distribution, which represent the kind of particles with majority in the distribution. The correlations of patterns representing other particles will usually remain below $c_{limit}^0$ and (if the number of majority particles are more than half of the number of all particles) do not contribute to the 3D distribution. When convergence is reached, the 3D intensity distribution will represent only the majority particles.

In the case when the number of the majority particles is smaller than half of the number of all particles *N*, (this can happen only with more than two kinds of particles) the above selection rule must be slightly modified. The median is be replaced with a selection of the best (having the largest correlations) *n* patterns, *n*/*N* < 0.5.

**Results and Discussion**

First we test the ability of our method to separate diffraction images of cytochrome and lysozyme molecules. They have similar size, but completely different structure. The structures are known from conventional X-ray diffraction measurements and are shown in Fig. 1a and b. A mixture of diffraction images of 12000 cytochrome and 8000 lysozyme molecules in random orientations were calculated. Poisson noise corresponding to an X-ray intensity of $2 \cdot 10^{13}$ photons/pulse focused to a 100nm x 100nm spot was added. The important parameters of the simulated patterns are listed in Table 1. For $c_{limit}^0$ the value of 0.32 was chosen. The CM method solved the orientation problem for the majority (cytochrome) molecules. The evolution of the distribution of the correlation values $c_m^{max}$ is shown in Fig. 2. The blue line shows the value of $c_{limit}$ during the iteration process. After convergence, the distribution of $c_m^{max}$ splits into two parts. The final distribution of $c_m^{max}$ is shown in Fig. 3. Since we know from our simulation that which diffraction pattern belongs to which molecule, as verification we can also plot the distribution of correlation separately for the two kinds of molecules. Almost all diffraction patterns in the higher correlation part above $c_{limit}$ belong to the cytochrome molecules and contribute to the 3D intensity distribution, while those below $c_{limit}$ belong to the lysozyme molecules, and do not contribute to it. Thus the resulting 3D intensity distribution represents only the structure of the cytochrome molecules. The angular deviation of the orientations of the patterns from the true orientations of the molecules was less than 2 degrees, corresponding to the distance between angular grid points (Tegze and Bortel, 2012). Further refinement of the orientations can reduce the deviation by an order of magnitude (Tegze and Bortel, 2012).

In our second study we test the ability of the method to distinguish between diffraction patterns of molecules with only small differences in their structure. For this study we have chosen pure lysozyme and its complex with arginine. Their respective structures are shown in Fig. 1b and c. Again, a mixture of diffraction images of 12000 lysozyme and 8000 Arg-lysozyme molecules in random orientations were calculated. It can be seen from Fig. 4 and 5 that the CM method can separate these molecules as well after 20 iterations. The convergence is slower than in the case of cytochrome-lysozyme mixture, due to the smaller differences between the structures of the molecules. The number of iterations necessary to solve the orientation problem for one component of a mixture depends not only on the intensity of the X-ray pulse, the number of patterns and the differences in structure but also on the relative number of the components. For example, when mixing 16000 lysozyme patterns with 4000 Arg-lysozyme ones (but keeping all other parameters the same), the CM algorithm found the orientations of the majority molecules after 7 iterations.

In the third study, equal number (12000) of diffraction patterns of cytochrome, lysozyme and Arg-lysozyme were mixed. Since each component is less than 50% of the mixture, the median in the expression of $c_{limit}$ is replaced by the best 30% of the patterns. Starting again from a random 3D intensity distribution, after 26 iterations the CM method finds the orientations of the diffraction patterns of one component, which happens to be the cytochrome (Fig. 6 and 7a). The low-correlation peak in Fig. 7a now contains a mixture of lysozyme and Arg-lysozyme. Now we can remove the patterns with correlation above $c_{limit}$, and continue with the iteration. After further 30 iterations, the orientations of the Arg-lysozyme patterns are found (Fig. 6 and 7b). After removing again the patterns with correlation above $c_{limit}$ and continuing the iteration process, the orientation problem of the remaining component (lysozyme) can also be solved (Fig 6 and 7c). Note that since there are only small differences in structure between lysozyme and Arg-lysozyme, convergence was reached after a few iterations.

**Conclusions**

In single particle diffraction experiments the particle beam may contain more than one kind of particle. We have shown that the correlation maximization algorithm can select one kind of particle out of a mixture of two or more components based on their noisy diffraction patterns without any prior knowledge of their structure or concentration in the mixture. The method can find the orientation of these particles with great accuracy making possible the compilation of a consistent 3D diffraction intensity distribution and subsequent solution of the structure. Separation of the diffraction patterns of particles with small structural differences was also possible.

**Acknowledgments**

We are grateful to G. Oszlányi for discussions. This work was supported by the Hungarian OTKA grants K81348 and NK105691 and by the Bolyai János Scholarship of the Hungarian Academy of Sciences to GB.

**Figures**

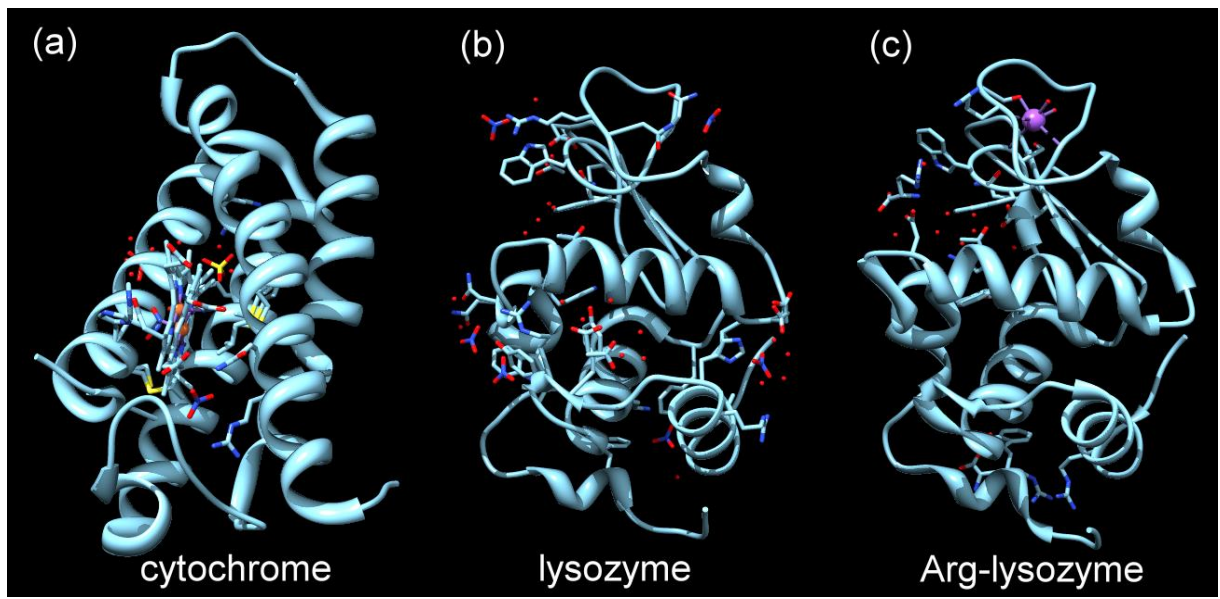

Figure 1. Structure of the molecules used in this study: cytochrome (a), lysozyme (b) and arginine-lysozyme complex (c).

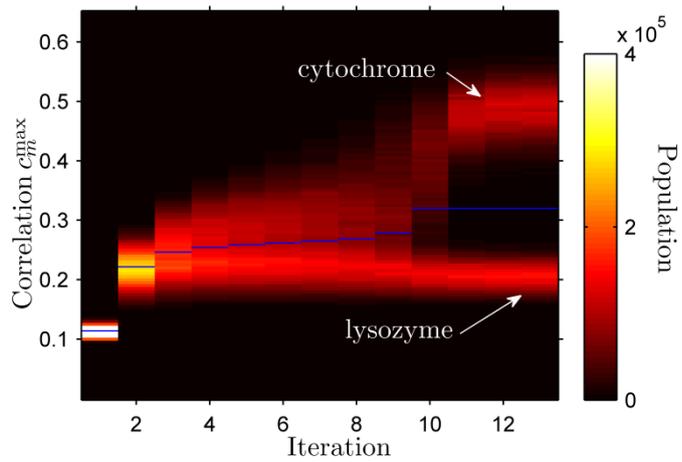

Figure 2. Evolution of the distribution of correlation maxima for the 3:2 mixture of cytochrome and lysozyme molecules. The blue line indicates the value of $c_{limit}$.

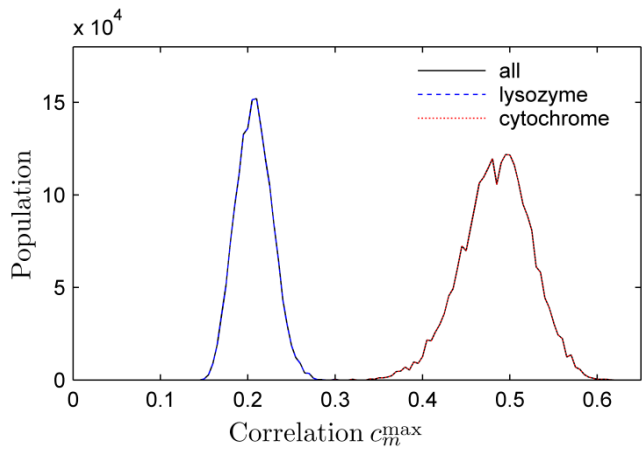

Figure 3. Distribution of correlation maxima for the 3:2 mixture of cytochrome and lysozyme molecules after convergence. The red (dotted) and blue (dashed) lines indicate the contribution of patterns of cytochrome and lysozyme, respectively. Since the separation is almost perfect, the black (solid) line indicating the full distribution is overlaid by the red (dotted) and blue (dashed) lines.

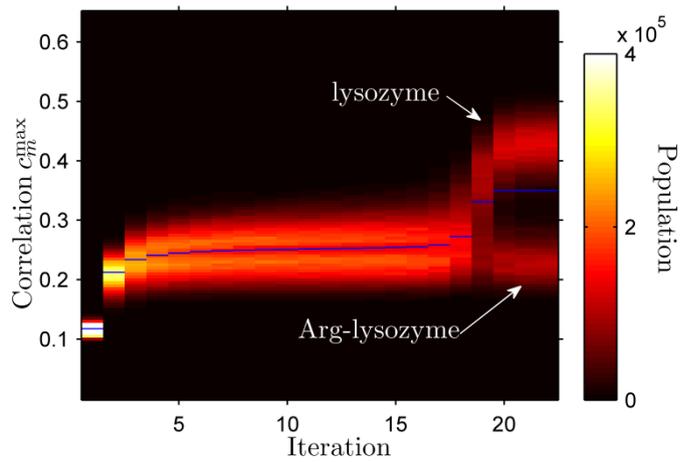

Figure 4. Evolution of the distribution of correlation maxima for the 3:2 mixture of lysozyme and Arg-lysozyme molecules. The blue line indicates the value of $c_{limit}$.

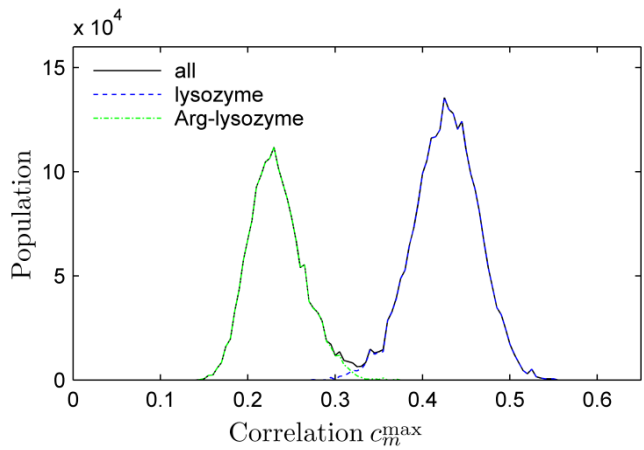

Figure 5. Distribution of correlation maxima for the 3:2 mixture of lysozyme and Arg-lysozyme molecules after convergence. The black (solid) line indicates the full distribution and the blue (dashed) and green (dash-dotted) lines indicate the contribution of patterns of lysozyme and Arg-lysozyme molecules, respectively.

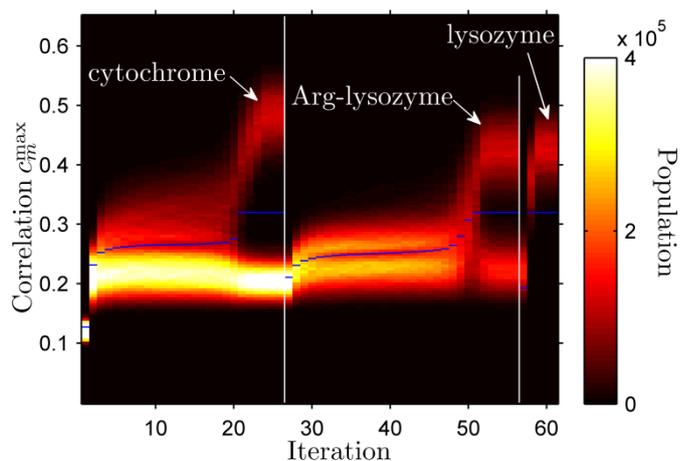

Figure 6. Evolution of the distribution of correlation maxima for the 1:1:1 mixture of cytochrome, lysozyme and Arg-lysozyme molecules. Patterns of already oriented components were removed from the mixture after iterations 26 and 56. The blue line indicates the value of $c_{limit}$.

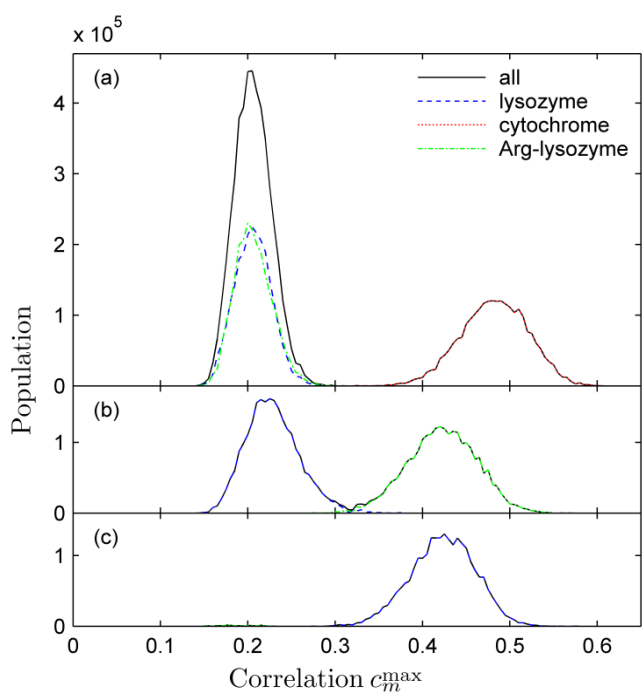

Figure 7. Distribution of correlation maxima for the 1:1:1 mixture of cytochrome, lysozyme and Arg-lysozyme molecules after iterations 26 (a), 56 (b) and 61 (c). The black (solid) line indicates the full distribution and the red (dotted), blue (dashed) and green (dash-dotted) lines indicate the contribution of patterns of cytochrome, lysozyme and Arg-lysozyme molecules, respectively.

**Tables**

Table 1. Parameters of simulated diffraction patterns and grids.

| | lysozyme | Arg-lysozyme | cytochrome |
|---|---|---|---|
| Diameter of the molecule, $D$ | 51.2 Å | 53.5 Å | 56.3 Å |
| Wavelength of the radiation, $\lambda$ | | 1 Å | |
| Pulse fluence | | $2\cdot10^{13}$ photons/(0.1μm×0.1μm) | |
| Minimum scattering angle, $\vartheta_{min}$ | | 1° | |
| Maximum scattering angle, $\vartheta_{max}$ | | 24° | |
| Pattern pixel size (polar × azimuthal), $\Delta\vartheta \times \Delta\varphi$ | | 1°× 2° | |
| Average total photon counts in pattern | 1702 | 1563 | 2149 |
| Average photon counts in outer pixels | 0.105 | 0.100 | 0.123 |
| Average photon counts in outer Shannon–Nyquist pixels (Huldt et al., 2003) of size $(\lambda/2D)^2$ | 0.041 | 0.035 | 0.039 |
| Number ($N_R$) and spacing of grid points in the $(\Theta,\Phi)$ orientation subspace (Tegze and Bortel, 2012) | | 5292 ~3° | |
| Number of grid points (voxels) in the reciprocal space, $N_I$ | | $129^3 = 2{,}146{,}689$ | |